\documentstyle[fleqn,epsfig]{article}
\begin{document}
\begin{flushright}
NORDITA -- 94/60 P\\
hep-ph/9411248
\end{flushright}
\vspace*{1cm}
%
%
\newcommand{\beq}{\begin{equation}}
\newcommand{\eeq}{\end{equation}}
\newcommand{\beqa}{\begin{eqnarray}}
\newcommand{\eeqa}{\end{eqnarray}}
\newcommand{\nn}{\nonumber}

\newcommand{\dd}{{\rm d}}
\newcommand{\mgluino}{m_{\tilde g}}
\newcommand{\msquark}{m_{\tilde q}}
\newcommand{\gluino}{\tilde g}
\newcommand{\squark}{\tilde q}
\newcommand{\alphas}{\alpha_{\rm s}}
\newcommand{\mZ}{m_Z}
\newcommand{\thW}{\theta_{\rm W}}
\newcommand{\GeV}{\mbox{{\rm GeV}}}

%
%
\newcommand{\hc}{\mbox{{\rm h.c.}}}
\newcommand{\Tsp}{\mbox{\scriptsize T}}
\newcommand{\mW}{m_{{ W}}}
\newcommand{\Md}{m_{{ d}}}
\newcommand{\Ad}{A_{{ d}}}
\newcommand{\sdq}{\tilde{{ d}}}
\newcommand{\Mu}{m_{{ u}}}
\newcommand{\Au}{A_{{ u}}}
\newcommand{\suq}{\tilde{{ u}}}
\newcommand{\suqa}{\tilde{{ u}}_{1}}
\newcommand{\suqb}{\tilde{{ u}}_{2}}
\newcommand{\sdqa}{\tilde{{ d}}_{1}}
\newcommand{\sdqb}{\tilde{{ d}}_{2}}

\newcommand{\MsQU}{\wtilde{3}{0.8}{M}_{\hspace*{-1mm}U}}
\newcommand{\Msu}{\wtilde{3}{0.2}{m}_{U}}
\newcommand{\Msd}{\wtilde{3}{0.2}{m}_{\!D}}
\newcommand{\Dis}[1]{$\displaystyle #1$}
\newcommand{\eps}{\epsilon}
\newcommand{\wtilde}[3]{\settowidth{\ltT}{\Dis{#3}}
\makebox[\ltT]{$\rule{#2\mmh}{0mm}
\widetilde{\makebox[#1\mm]{\Dis{#3\rule{#2\mm}{0mm}}}}$}}
\newcommand{\mgli}{m_{\tilde{g}}}
\newcommand{\gli}{\psi_{g}}
\newcommand{\ogli}{\overline{\psi}_{g}}

\newlength{\ltT}
\newlength{\mmh}
\setlength{\mmh}{0.5mm}
\newlength{\mm}
\setlength{\mm}{1mm}

\newcommand{\Glue}{\wtilde{3}{0.8}{G}}
\def\Order{{\cal O}}
\begin{center}
{\large \bf LIGHT-GLUINO PRODUCTION AT LEP}\footnote{To appear in:
{\it Proceedings of IX International Workshop: High Energy
Physics and Quantum Field Theory}, Zveni\-gorod, Russia,
September 16--22, 1994} \\

\vspace{4mm}
Bjarte Kileng \\
NORDITA, Blegdamsvej 17, DK-2100 Copenhagen \O, Denmark\\
\vspace{4mm}
Per Osland \\
University of Bergen, All\'egt.~55, N-5007 Bergen, Norway
\end{center}

\vspace{6mm}  
If gluinos are light, they will be
produced in electron-positron annihilation at LEP, not only by
radiation in pairs off quarks and antiquarks, but also without
accompanying quark and antiquark jets.
We here discuss the latter process,
pair production of gluinos,
in a model with soft supersymmetry breaking,
allowing for mixing between the squarks.
In much of the parameter space of the Minimal Supersymmetric Model (MSSM)
the cross section corresponds
to a $Z$ branching ratio above $10^{-5}$, even up to $10^{-4}$.
A non-observation of gluinos at this level
restricts the allowed MSSM parameter space.

\renewcommand{\theequation}{\thesection.\arabic{equation}}
\section{Introduction}
\label{sec:intro}
\setcounter{equation}{0}

Moderately heavy gluinos, decaying to a photino,
would be produced in hadronic collisions, and lead to
events with missing energy (the photinos).
Recent searches for gluinos by the CDF Collaboration have
thus established a lower mass bound of the order of
140~GeV/${\rm c}^2$ \cite{CDF}.
This bound depends on the assumed decay mode of the gluino,
it is valid for the case of direct decay to the lightest
supersymmetric particle, $\gluino\to q \bar q \tilde\chi$.
The analysis is insensitive to light gluinos,
$\mgluino\le\Order(40\ \GeV/{\rm c}^2)$.
However, various other experiments, in particular those at the CERN SPS
\cite{UA1}
exclude, for short-lived gluinos, most of the region below
40~GeV/${\rm c}^2$, except for a narrow range around
a few GeV/c${}^2$ \cite{PDG}.

The existence of this low-mass gluino window has been given
some attention recently \cite{UA1,PDG},
and it is argued that data on $\alphas(\mZ)$ favour
a light gluino \cite{Jezabek}.
It is easy to imagine that the gluino mass is induced radiatively,
in which case it would naturally be light
\cite{FarMas}.
\begin{figure}[thb]
\centerline{\psfig{figure=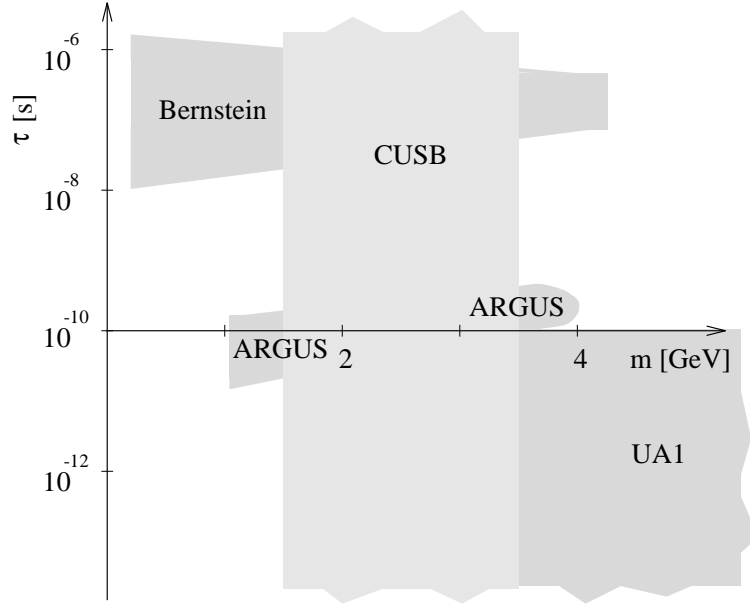,width=10cm}}  
\vspace*{1mm}
\caption{\label{plfun1}Excluded regions of gluino mass and lifetime.}
\end{figure}
The importance of searching for light gluinos has repeatedly been stressed
by Farrar \cite{Farrar,Farrar94}.
Clearly, if the gluino is very light, it should be produced at LEP,
either by radiation in pairs off a quark \cite{Farrar,Campbell1},
or in pairs via
the triangle diagram \cite{Nelson,Kane,Campbell2}.
In the former case, there is some uncertainly about how difficult
it would be to isolate the final four-jet state,
because of the QCD background \cite{Hultqvist}.
For the latter mechanism, the cross section
has recently been re-evaluated, taking into account a heavy top
quark mass and the effects of chiral mixing \cite{KilOsl}.

Light gluinos have actually been ruled out for some ranges of mass and
lifetime by a variety of experiments. A recent survey has been compiled
by Farrar \cite{Farrar94} and is schematically reproduced as Fig.~1.
The CUSB experiment \cite{CUSB}
was sensitive to gluinos of any lifetime longer than the hadronization
time scale, while others, like ARGUS \cite{ARGUS} and UA1 \cite{UA1}
were only sensitive to gluinos decaying within the detector.
The experiment by Bernstein et al.\ \cite{Bernstein}
searched for single charged particles resulting from the decays
of neutral particles of lifetimes between $10^{-8}$
and $2\cdot10^{-6}$~sec.
Their result has been interpreted in terms of excluded gluino masses
and lifetimes by Farrar \cite{Farrar94}.

Such plots are sometimes given in terms of squark masses instead
of lifetimes. If the gluino decays to a {\it massless} photino,
then the lifetime is related to the lightest relevant squark mass by
\cite{Dawson}
\beq
\tau\simeq4\cdot10^{-8}\mbox{ sec}
\left[{\msquark\over 1 \mbox{ TeV/c$^2$}}\right]^4
\left[{1 \mbox{ GeV/c$^2$} \over \mgluino}\right]^5 \,.
\eeq

The model considered in ref.~\cite{KilOsl} is in part given by
the (soft) supersymmetry breaking part of the Lagrangian,
which is given in terms of component fields as \cite{Kileng}
\begin{eqnarray}
\label{EQU:Lagrangefour}
{\cal L}_{\mbox{{\scriptsize Soft}}}
& = & \Biggl\{
      \frac{g\Md\Ad}{\sqrt{2}\;\mW\cos\beta}Q^{\Tsp}\eps H_{1}\sdq^{R}
    - \frac{g\Mu\Au}{\sqrt{2}\;\mW\sin\beta}Q^{\Tsp}\eps H_{2}\suq^{R}
    + \hc \Biggr\} \nonumber \\
& & - \MsQU^{2}Q^{\dagger}Q - \Msu^{2}\suq^{R\dagger}\suq^{R}
    - \Msd^{2}\sdq^{R\dagger}\sdq^{R}
    + \frac{\mgli}{2} \sum_{a=1}^{8} \left({\gli}_{a}{\gli}_{a}
    + {\ogli}_{a}{\ogli}_{a}\right)   \hspace*{1mm}.
\end{eqnarray}
Subscripts $u$ (or $U$) and $d$ (or $D$) refer generically to up and
down-type quarks.

The gluino mass is given explicitly by $\mgluino$, whereas squark
masses depend not only on the mass parameters $\MsQU$,
$\Msu$ and $\Msd$, but also on $m_u$, $m_d$, $m_Z$, $m_W$,
$A_u$, $A_d$, $\mu$ and $\beta$.
Here, $\mu$ is the coupling between the two Higgs doublets, and
$\tan\beta$ the ratio of the two Higgs vacuum expectation values.
The somewhat lengthy mass formulas are \cite{Kileng,Brignole}:
\beqa
m^2_{\tilde u\,\{1,2\}}&=& m_u^2+\frac{1}{2}\left(\MsQU^2+\Msu^{2}\right)
+\frac{\mZ^2}{4}\cos(2\beta) \nn \\
& &
\pm\sqrt{\left[\frac{1}{2}\left(\MsQU^2-\Msu^{2}\right)
+\frac{1}{2}\left(\frac{4}{3}\mW^2-\frac{5}{6}\mZ^2\right)\cos(2\beta)\right]^2
+m_u^2\left|A_u+\mu\cot\beta\right|^2}
\label{EQU:massesu} \\
m^2_{\tilde d\,\{1,2\}}&=& m_d^2+\frac{1}{2}\left(\MsQU^2+\Msd^{2}\right)
-\frac{\mZ^2}{4}\cos(2\beta) \nn \\
& &
\pm\sqrt{\left[\frac{1}{2}\left(\MsQU^2-\Msd^{2}\right)
+\frac{1}{2}\left(\frac{-2}{3}\mW^2+\frac{1}{6}\mZ^2\right)\cos(2\beta)\right]^2
+m_d^2\left|A_d+\mu\tan\beta\right|^2}
\label{EQU:massesd}
\eeqa

It should be noted that the above Lagrangian represents a model
which is different from the recently considered
``constrained" models based on Grand Unification and supergravity
\cite{RGR},
the gluino mass is here not tied to the other gaugino masses.
\begin{figure}[t]
\begin{center}
\setlength{\unitlength}{1cm}
\begin{picture}(16,8.5)
\put(0.8,-2){\mbox{\epsfysize=17cm\epsffile{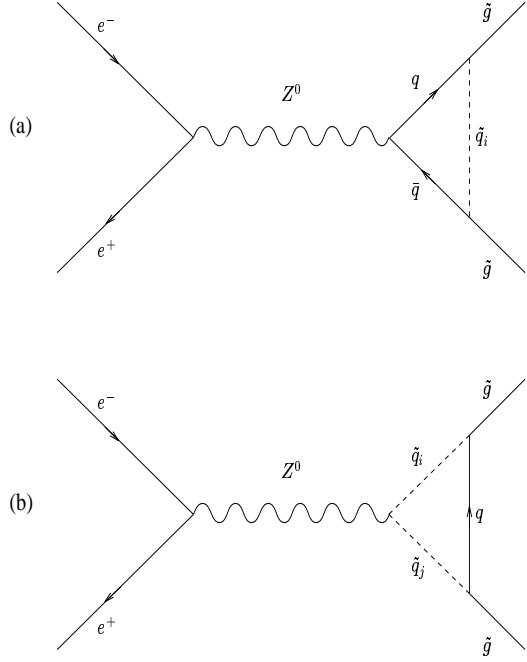}}}
\end{picture}
\vspace*{-8mm}
\caption{\label{plfun2}The two classes of Feynman diagrams
for $e^{+}\,e^{-} \rightarrow \tilde{g}\,\tilde{g}$.}
\end{center}
\end{figure}

\section{The $e^+e^-\to\gluino\gluino$ Cross Section}
\setcounter{equation}{0}
\label{sec:ampl}
In the decay of the $Z$, or more generally in electron-positron
annihilation, the pair production of gluinos can proceed via the two generic
diagrams $(a)$ and $(b)$ of Fig.~2, where the internal lines
of the triangles are quarks and squarks.

The amplitude for
\beq
e^+e^- \to \gluino\gluino
\eeq
will be proportional to the gluino current which can be written as
a sum of contributions from the different quark flavours
associated with the triangle diagrams,
with the $u$-quark contribution
\beqa
\label{EQU:glueu}
\Glue^\mu_u
&=&
\left(\Glue^\mu_{uu1} + \Glue^\mu_{uu2}\right)
+ \left({\Glue}_{11u}^{\mu} + {\Glue}_{22u}^{\mu}
+       {\Glue}_{12u}^{\mu} + {\Glue}_{21u}^{\mu}\right)
+       \mbox{crossed terms} \, .
\eeqa
The different labels refer to the quark and squark propagators of the
triangle diagram.


We find that the cross section is proportional to the square of the sum of
two partial amplitudes,
corresponding to the contributions of the two diagrams (a) and (b).
This is possible, since
there is essentially only one invariant amplitude \cite{Nelson}.
The integrated cross section thus takes the form
\beq
\label{EQU:sigma}
\sigma
= \frac{g^{2}\pi^{3}\left(g_{V}^{2}+g_{A}^{2}\right)
\left(\sqrt{E^2-\mgli^{2}\,}\,\right)^{3}}{12E\cos^{2}\thW
\left[ \left(s - \mZ^{2} \right)^2
+ \Gamma_Z^{2}\mZ^{2} \right]} \;
\left|\sum({\cal A}_a+{\cal A}_b)\right|^2,
\eeq
with $E$ the beam energy and
the sum running over quark flavours $q$.
The two partial amplitudes correspond to diagrams (a) and (b) and are
given in ref.\ \cite{KilOsl}.

Actually, since there is only one invariant
amplitude, whose structure is determined by the fact that it
describes the annihilation of two massless fermions to a pair
of self-conjugate fermions \cite{Nelson}, the angular distribution
is given by the familiar $1+\cos^2\theta$.
%

\section{Conditions for Vanishing Cross Section}
\setcounter{equation}{0}
\label{:vanish}

In order to better understand what is required for the cross
section to be large, let us first state the conditions that must
be satisfied in order for it to vanish.

The gluino pair production cross section would {\it vanish}
if the following
{\it conditions were both satisfied} \cite{Campbell2}
\begin{itemize}
\begin{enumerate}
\item mass degeneracy in each quark isospin doublet,
$m_d=m_u$ (this is violated),

\item mass degeneracy in each squark isospin doublet, i.e.,
$m_{{\tilde d}_1}=m_{{\tilde d}_2}
=m_{{\tilde u}_1}=m_{{\tilde u}_2}$, for each generation.

\end{enumerate}
\end{itemize}
For comparison, in the case of no axial coupling to the $Z$,
i.e., in the QED limit,
the cross section would {\it vanish} if there is \cite{Nelson}

\begin{itemize}
\item mass degeneracy in each squark chiral doublet,
i.e., $m_{\tilde u1}=m_{\tilde u2}$, and $m_{\tilde d1}=m_{\tilde d2}$
for each generation.
This condition is less strong than item~(2) above.
\end{itemize}

The magnitude of the cross section will depend on how strongly
these conditions (1) and (2) are violated.
Especially for the third generation, item~(1) is violated.
This is generally believed to imply that the squark isospin
doublets are not degenerated either.
However, in a consistent MSSM, the squark masses
can not be specified as free parameters, they emerge as dependent
on the more fundamental parameters of the Lagrangian.

One may ask whether it is possible for all squark masses to be degenerate.
{}From eqs.~(\ref{EQU:massesu}) and (\ref{EQU:massesd}), this is seen to
require
\beq
A_u+\mu\cot\beta=A_d+\mu\tan\beta=0
\eeq
and (invoking $\mW^2=\cos^2\thW\mZ^2$)
\beq
\cos(2\beta)=-\frac{m_u^2-m_d^2}{\mW^2} \,.
\eeq
The last condition clearly cannot be satisfied for a realistic
top quark mass,
so we conclude that there will inevitably be a contribution
to the gluino pair-production cross section from the
third generation.
\section{Results}
\setcounter{equation}{0}
\label{:results}

For the purpose of developing some intuition for how large
the gluino pair production cross section would be at LEP,
we show in Fig.~3 the ratio
\beq
R=\frac{\sigma(e^+e^-\to\gluino\gluino)}{\sigma(e^+e^-\to\mu^+\mu^-)}
\eeq
vs.\ maximal squark mass splitting $\delta \msquark$.
\begin{figure}[thb]
\begin{center}
\setlength{\unitlength}{1cm}
\begin{picture}(16,10)
\put(0.2,-4.4){\mbox{\epsfysize=20cm\epsffile{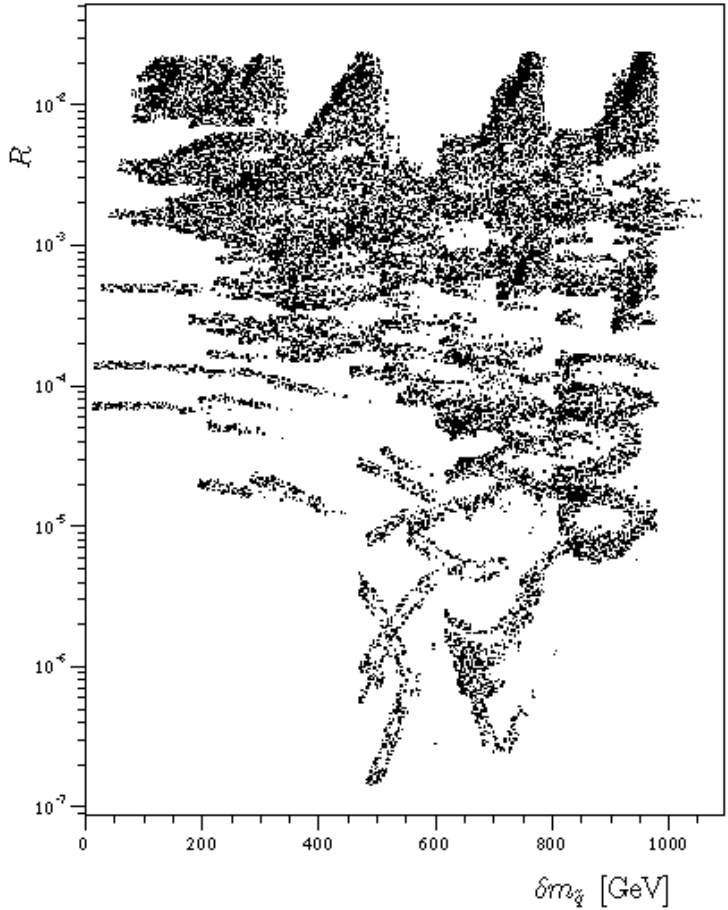}}}
\end{picture}
\vspace*{0mm}
\caption{Cross section ratios $R=\sigma(e^+e^-\to\gluino\gluino)/
\sigma(e^+e^-\to\mu^+\mu^-)$ at the $Z$ resonance.
The figure shows the result of a scan of parameter space,
against the {\it largest} resulting
squark mass difference.}
\end{center}
\vspace*{-2mm}
\end{figure}
The plot is based on a scan of the MSSM parameter space \cite{KilOsl},
for gluino, bottom and top quark masses given by
$\mgluino=3.5~\GeV/{\rm c}^2$, $m_b=4.8~\GeV/{\rm c}^2$,
and $m_t=170~\GeV/{\rm c}^2$.
All encountered cross section ratios are represented by dots
in this scatter plot.
The horizontal axis gives the largest
resulting squark mass difference.
The cross section ratios are seen to be typically
between $10^{-4}$ and $10^{-2}$.
The $Z$ branching ratio is obtained upon multiplying by 3.3\%.
Parameter sets that lead to
any one of the squarks being light, $\msquark<45~\GeV/{\rm c}^2$,
are left out, since such light squarks would have been detected at LEP
\cite{LEPsquark}.

The band structures are ascribed
to the discreteness of the sampling, as well as the rather complex dependence
the cross section has on the many parameters.
If the squark masses are taken
as free parameters, then the cross section displays narrow valleys
in the space of squark masses.
Some of these valleys are related to the regions of
vanishing cross sections quoted in Sect.~3,
but there are also regions of low cross section not directly
related to the conditions of Sect.~3, as illustrated in Fig.~4.

The value for the gluino mass, $\mgluino=3.5~\GeV/{\rm c}^2$, has been
chosen as representative of the ``light-gluino window".
Actually, the cross section has only a very weak dependence
on the gluino mass, as long as it is well below
the kinematical threshold \cite{Kilengthesis}.

\begin{figure}[thb]
\begin{center}
\setlength{\unitlength}{1cm}
\begin{picture}(16,10)
\put(0.0,-4.2){\mbox{\epsfysize=20cm\epsffile{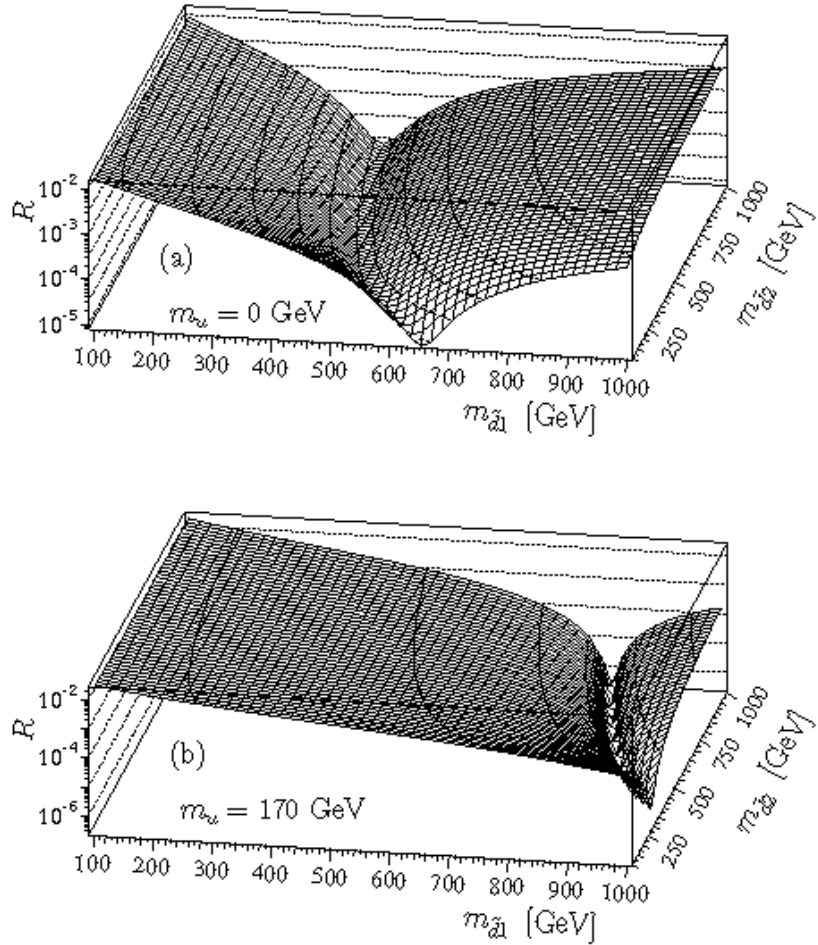}}}
\end{picture}
\vspace*{3mm}
\caption{Cross section ratio $R$ as a function of squark masses
$m_{\tilde d1}$ and $m_{\tilde d2}$, for two values of $m_u$, and
with $m_{\tilde u1}=1$~TeV/c${}^2$ and $m_{\tilde u2}=100$~GeV/c${}^2$.}
\end{center}
\vspace*{-2mm}
\end{figure}

\section{Discussion}
\setcounter{equation}{0}
\label{:discussion}
Even though the cross section for producing light gluinos could
be rather large at LEP, their actual discovery might be difficult.
First of all, we note that their contribution to the total $Z$ width,
2490~MeV$\cdot R$, would lead to a small surplus of two-jet events,
as compared with three-jet events, and thus a minute reduction
of $\alphas$.

For the purpose of discussing the signatures of gluino jets,
let us consider the following cases separately:
(i) the gluinos are unstable and decay within the detector,
$\tau_{\gluino}\le\Order(10^{-9}\sec)$ (corresponding to
a squark mass of $\msquark\le2$~TeV/c${}^2$), or
(ii) the gluinos are long-lived or stable, and do not decay
inside the detector.

If the gluinos do not decay within the detector,
their discovery would be very difficult.  They would fragment to jets
consisting of ordinary hadrons and a leading $R$-hadron.
The lightest $R$-hadron is believed to be
the gluinoball, $R_0$, consisting of $\gluino g$
\cite{Farrar,Farrar94}.
This will presumably interact a few times
in the calorimeter, depositing a major fraction of its energy.
Such events would therefore be hard or impossible to distinguish
from ordinary $q\bar q$ events.

The more hopeful situation is when the gluinos decay within the
detector, such that the resulting events will have missing energy due
to the escaping neutralino.  This is the standard SUSY signal.
The most serious background will presumably be from $b\bar b$ events,
where some energy is carried away by neutrinos.
But the semileptonic decay of the $b$ also results in a charged lepton,
which in principle should distinguish these events from the gluino
events.

Another kind of background would be $q\bar q$ events in which
not all the energy of the neutrals is detected.
Again, this appears to be a less serious problem, but a dedicated
Monte Carlo study might be necessary in order to fully understand
these backgrounds.

In summary, the pair production of light gluinos, without accompanying
quark jets, is in $Z$ decay large enough to be measurable in
much of the MSSM parameter space, and should therefore
be searched for vigorously.

\medskip
It is a pleasure to thank F. Cuypers, T. Medcalf, F. Richard,
and A. Sopczak for useful comments.
We are also grateful to the Organizers of the Zvenigorod Workshop,
in particular Professor V. Savrin,
for creating a very stimulating and pleasant atmosphere during the meeting.
This research has been supported by the Research Council of Norway.

\vspace{12mm}

\end{document}